\def\aa{{A\&A}}
\def\aas{{A\&AS}}
\def\aj{{AJ}}
\def\annrev{{ARA\&A}}
\def\apj{{ApJ}}
\def\apjs{{ApJS}}
\def\baas{{BAAS}}
\def\mnras{{MNRAS}}
\def\nat{{Nature}}
\def\pasp{{PASP}}

\documentclass[cup5b]{caps}
\usepackage{graphicx}
\usepackage{amssymb}
\usepackage{ociwsymp4}
\HeadText{M. Loewenstein}

\def\plottwo#1#2{\centering \leavevmode
\includegraphics[width=.45\columnwidth]{#1} \hfil
\includegraphics[width=.45\columnwidth]{#2}}

\def\plottwo90#1#2{\centering \leavevmode
\includegraphics[width=.45\columnwidth,angle=90]{#1} \hfil
\includegraphics[width=.45\columnwidth,angle=90]{#2}}

\begin{document}
\pagenumbering{arabic}

\author[]{MICHAEL LOEWENSTEIN\\NASA/Goddard Space Flight Center and
University of Maryland}

\chapter{Chemical Composition of the Intracluster Medium \\ }

\begin{abstract}
Clusters of galaxies are massive enough to be considered
representative samples of the Universe, and to retain all of the heavy
elements synthesized in their constituent stars. Since most of these
metals reside in hot plasma, X-ray spectroscopy of clusters provides a
unique and fundamental tool for studying chemical evolution. I review
the current observational status of X-ray measurements of the chemical
composition of the intracluster medium, and its interpretation in the
context of the nature and history of star and galaxy formation
processes in the Universe. I provide brief historical and cosmological
contexts, an overview of results from the mature {\it ASCA}
observatory database, and new results from the {\it Chandra} and {\it
XMM-Newton} X-ray observatories. I conclude with a summary of
important points and promising future directions in this rapidly
growing field.
\end{abstract}

\section{Introduction}

\subsection{Rich Clusters of Galaxies and their Cosmological Setting}

Rich clusters of galaxies, characterized optically as concentrations
of hundreds (or even thousands) of galaxies within a region spanning
several Mpc, are among the brightest sources of X-rays in the sky,
with luminosities of up to several $10^{45}$ erg s$^{-1}$. The hot
intracluster medium (ICM) filling the space between galaxies has an
average particle density $\langle n_{\rm ICM}\rangle\approx 10^{-3}$
cm$^{-3}$ and an electron temperature ranging from 20 to $>100$
million K. Interpreted as virial temperatures, these correspond to
masses of $\sim (1-20) \times 10^{14}$ $M_{\odot}$; rich clusters are
believed to be the largest gravitationally bound structures in the
Universe. They are also notable for their high fraction (typically
$\sim 75$\%) of member galaxies of early-type morphology, as compared
to galaxy groups or the field. Within the framework of large-scale
structure theory, rich clusters arise from the largest fluctuations in
the initial random field of density perturbations, and their
demographics are sensitive diagnostics of the cosmological world model
and the origin of structure (e.g., Schuecker et al. 2003). Rich galaxy
clusters are rare and (including their dark matter content) account
for less than 2\% of the critical density, $\rho_{\rm crit}$, characterizing a 
flat Universe.

Embedded in the ICM of rich clusters---where
$70\%-80\%$ of cluster metals reside---lies a uniquely accessible
fossil record of heavy element creation. To the extent that the
cluster galaxy stars where these metals were synthesized are
representative, measurement of ICM chemical abundances provides
constraints on nucleosynthesis---and, by extension, the epoch,
duration, rate, efficiency, and initial mass function (IMF) of star
formation---in the Universe. From this perspective it is useful to
take an inventory of clusters, and compare with the Universe as a
whole.

Consideration of recent results from the {\it Wilkinson Microwave
Anisotropy Probe} supports a standard cosmological model wherein, to
high precision, the average matter density totals $0.27\rho_{\rm
crit}$ in a flat Universe, and baryons amount to $0.044\rho_{\rm
crit}$ (Spergel et al. 2003). The estimate of Fukugita, Hogan, \&
Peebles (1998) of the total density in stars, $\sim 0.0035\rho_{\rm
crit}$, is corroborated by recent constraints based on the
extragalactic background light (Madau \& Pozzetti 2000), and the Two
Micron All-Sky Survey and Sloan Digital Sky Survey (Bell et
al. 2003). Since critical density corresponds to a mass-to-light ratio
($B$ band) $M/{L_B}\approx 1000$, the cluster matter inventory compares to
the Universe as a whole as indicated in Table 1.1.

\begin{table}
\caption{Mass-to-Light Ratios and Mass Fractions}
\begin{tabular}{cccc|cccc|c}
\hline\hline
Parameter &&&& Universe &&&& Clusters \\ \hline
$\langle{M_{\rm total}/{L_B}}\rangle$ &&&& 270   &&&& 300 \\
$\langle{M_{\rm stars}/{L_B}}\rangle$ &&&& 3.5   &&&& 4 \\
$\langle{M_{\rm gas}/{L_B}}\rangle$   &&&& 41    &&&& 35 \\
$f_{\rm baryon}$                      &&&& 0.17  &&&& 0.13 \\
$f_{\rm stars}$                       &&&& 0.013 &&&&0.013 \\
$f_{\rm gas}$                         &&&& 0.15  &&&& 0.12 \\
stars/gas                             &&&& 1/12  &&&& 1/9 \\
\hline\hline
\end{tabular}  
\label{table 1.1}
\end{table}

Deviations from the typical rich cluster values displayed in Table 1.1
are found for both the total mass-to-light ratio and the baryonic
contributions (Ettori \& Fabian 1999; Mohr, Mathiesen, \& Evrard 1999;
Bahcall \& Comerford 2002; Girardi et al. 2002; Lin, Mohr, \& Stanford
2003)---indicative of differences and uncertainties in assumptions,
method of calculation, and in extrapolation to the virial radius, as
well as possible cosmic variance. However, it is clear that, at least
to first order, observations are consistent with the theoretical
expectation (e.g., Evrard 1997) that these largest virialized
structures are ``fair samples'' of the Universe in terms of their mix
of stars, gas, and dark matter. (A corollary of this is that bulge
populations generally dominate the stellar mass budget in the field,
as well as in clusters.) While clusters were the first systems
identified with baryon budgets dominated by a reservoir of hot gas
(White et al. 1993), this is now believed to apply to the Universe as
a whole at past and present epochs (Dav\'e et al. 2001; Finoguenov,
Burkert, \& Bohringer 2003).

An important {\it caveat} is that, since rich clusters do represent
regions of largest initial overdensity, star formation may initiate at
higher redshift, proceed with enhanced efficiency, or be characterized
by an IMF skewed toward higher mass stars when
compared to more typical regions. If so, there is an opportunity to
search for possible variations in star formation with epoch or
environment, given a suitable class of objects for comparison.

The intergalactic medium in {\it groups} of galaxies may comprise one
such sample. Groups generally include $\sim 2-50$ member galaxies,
emit at X-ray luminosities $<10^{44}$ erg s$^{-1}$, and have electron
temperatures $<20$ million K, corresponding to mass scales up to $\sim
10^{14}$ $M_{\odot}$ (Mulchaey 2000). Groups present their own unique
interpretive challenges. They may not behave as closed boxes and
evidently display a spread in their mass inventories, metallicities,
and morphological mix of galaxies that reflect the theoretically
expected cosmic variance in formation epoch and evolution (Davis,
Mulchaey, \& Mushotzky 1999; Hwang et al. 1999). However, since
extending consideration to the poorest groups encompasses most of the
galaxies (and stars) in the Universe, it is crucial to study the
chemical composition of the intragroup, as well as the intracluster,
medium.

\subsection{Advantages of X-ray Wavelengths for Abundance Studies}

From both scientific and practical perspectives, X-ray spectroscopy is
uniquely well suited for studying the chemical composition of the
Universe. Most of the metals in the Universe are believed to reside in
the intergalactic medium; this is demonstrably true for rich
galaxy clusters (e.g., Finoguenov et al. 2003). The concordance in
mass breakdown between clusters and the Universe discussed above
implies that, because of their deep potential wells, clusters, unlike most, 
if not all, galaxies, are ``closed boxes'' in the
chemical evolution sense. Thus, modelers are provided with an unbiased
and complete set of abundances that enables extraction of robust
constraints on the stellar population responsible for metal enrichment.

For high temperatures, such as those found in the ICM, the shape of
the thermal continuum emission yields a direct measurement of the
electron temperature, and thus the ionization state. Complications
arising from depletion onto dust grains, optical depth effects, and
uncertain ionization corrections are minimal or absent. The energies
of K-shell ($\rightarrow n=1$) and/or L-shell ($\rightarrow n=2$)
transitions for {\it all} of the abundant elements synthesized after
the era of Big Bang nucleosynthesis lie at wavelengths accessible to
modern X-ray astronomical telescopes and instruments. The strongest
ICM emission lines arise from well-understood H- and He-like ions, and
line strengths are immediately converted into elemental abundances via
spectral fitting. Of course, the quality and usefulness of X-ray
spectra are limited by the available sensitivity and spectral
resolution. In the following sections, I detail how the rapid
progression in the capabilities of X-ray spectroscopy drives the
evolution of our understanding of intracluster enrichment and its
ultimate origin in primordial star formation in cluster galaxies. New
puzzles are revealed with every insight emerging from subsequent
generations of X-ray observatories, a situation that will surely
continue with future missions.

\section{Earliest Results on Cluster Enrichment}

The 6.7 keV He-like Fe K$\alpha$ emission line is the most easily
detectable feature in ICM X-ray spectra, because it arises from a 
high-oscillator strength transition in an abundant ion and lies in a
relatively isolated wavelength region. It was first detected in the
{\it Ariel V} spectrum of the Perseus cluster (Mitchell et al. 1976),
and {\it OSO-8} observations of the Perseus, Coma, and Virgo Clusters
(Serlemitsos et al. 1977). Spectroscopic analysis of $\sim 30$
clusters (many observed with the {\it HEAO-1 A-2} satellite) revealed the
ubiquity of this feature at a strength generally indicating an ICM Fe
abundance of one-third to one-half solar (Mushotzky 1984). These early
results demonstrated the origin of cluster X-ray emission as a thermal
primordial plasma enriched by material processed in stars and ejected
in galactic winds. Although not immediately realized, this provided
the first indications of the prodigious magnitude of galactic
outflows, since the measured amount of ICM Fe was of the same order as
the sum of the Fe in all of the stars in all of the cluster galaxies.

Prior to 1993, Fe was the only element accurately measured in a large
number of clusters, although some pioneering measurements of O, Mg,
Si, and S features were obtained with the Solid State Spectrometer and
Focal Plane Crystal Spectrometer aboard the {\it Einstein} X-ray
Observatory (see Sarazin 1988 and references therein for these
results, as well as a complete survey of the field prior to the the
launch of the {\it ROSAT} and {\it Advanced Satellite for Cosmology and 
Astrophysics (ASCA)} X-ray observatories). Given
the ambiguous origins of Fe, known to be synthesized in comparably
large quantities by both Type Ia and Type II supernovae (henceforth,
SNe~Ia and SNe~II) integrated over the history of the Milky Way, this
represented a serious impediment to interpreting the data in terms of
star formation history. Fortunately, the launch of {\it ASCA} provided the 
means to utilize the ICM as a repository of information on historical star
formation and element creation.

\section{ICM Abundances from the {\it ASCA} Database}

With the combination of modest imaging capability, good spectral
resolution, and large collecting area over a broad X-ray bandpass
realized by its four telescope/detector pairs, the {\it ASCA} X-ray
Observatory ushered in a new era of X-ray spectroscopy of
astrophysical plasmas and revolutionized cluster elemental abundance
studies (Arnaud et al. 1994; Bautz et al. 1994; Fukazawa et al. 1994).
The clusters with accurate Fe abundance determinations cover a
range of redshifts out to $z\approx 0.5$. This enabled the first
investigation of the evolution of [Fe/H], as needed to understand the
epoch and mechanism of enrichment (Mushotzky \& Loewenstein
1997). The first results on gradients in ICM metallicity and abundance
ratios were obtained (see \S 1.5), and the first significant sample of
$\alpha$-element abundances were measured with {\it ASCA}. Well over
400 observations of galaxy clusters and groups were made over the
1993--2000 duration of the {\it ASCA} mission. Because the {\it ASCA}
data archive is now complete, and a comparable database based on
observations with the current generation of X-ray observatories is
many years away, an assessment of {\it ASCA} ICM enrichment results is
timely.

The {\it ASCA} Cluster Catalog (ACC) project (Horner et al. 2004),
initiated by K. Gendreau (NASA/GSFC), is the subject of the University
of Maryland Ph.D. theses of D. Horner and W. Baumgartner, and includes
contributions from R. Mushotzky, C. Scharf, and the author. Its
overarching goal is to perform uniform and semi-automated spectral
analysis on the integrated X-ray emission from every target in the
{\it ASCA} cluster observational database, utilizing the most current
processing and calibration. {\it ASCA} observed 434 galaxy clusters
and groups in 564 pointings. Among these, 273 clusters are deemed suitable for
spectroscopy, superseding previous samples (e.g., White 2000) by more
than a factor of 2.

\begin{figure*}
\centering
\includegraphics[clip,height=1.25\columnwidth,angle=0]{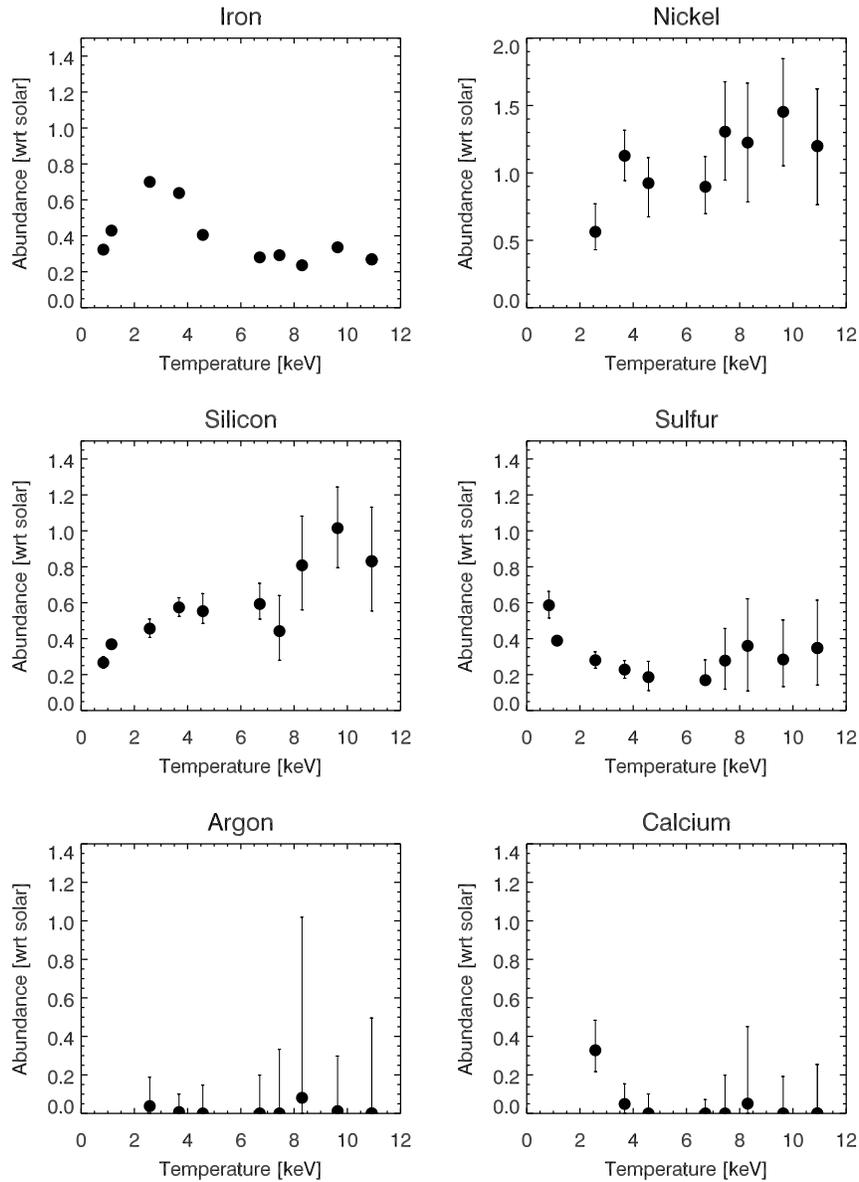}
\vskip 0pt \caption{
Average abundances with respect to the solar abundances of
Grevesse \& Sauval (1998) as a function of temperature. 90\%
confidence errors are shown. (From Baumgartner et al. 2004.)}
\label{av_abund}
\end{figure*}

\subsection{Fe Abundance Trends}

The Fe abundance shows a remarkable uniformity in rich clusters, with
a distribution that sharply peaks at $\sim 0.4$ solar (where solar Fe
abundance is defined such that log (Fe/H) + 12 = 7.5; Grevesse \& Sauval
1998). The Fe abundance tends to be higher in clusters with the
highest central densities and shortest cooling times (see also Allen
\& Fabian 1998). It is unclear from these data alone whether this is
an artifact of the presence of Fe gradients: since the X-ray
emissivity is proportional to the gas density squared, X-ray spectra
of more concentrated ICM are disproportionately weighted toward the
more abundant central regions. There are also systematic trends with
electron temperature, with Fe tending to be overabundant with respect
to the average defined by the hottest clusters, for clusters with ICM
temperatures 2 keV $<\,kT\,<\,4$ keV, and underabundant for $kT<2$ keV.

{\it ASCA} data revealed evidence for a {\it lack} of evolution in Fe
abundance out to redshift 0.4 (Mushotzky \& Loewenstein 1997; Rizza et
al. 1998; Matsumoto et al. 2000, 2001) and no
evidence of a decline at higher redshift (Donahue et al. 1999). {\it
XMM-Newton} and {\it Chandra} measurements are further expanding the
redshift range of precise cluster Fe abundance measurements (Jeltema
et al. 2001; Arnaud et al. 2002; Maughan et al. 2003; Tozzi et
al. 2003). On average, rich cluster Fe abundances are invariant out to
$z\approx 0.8$, and show no sign of decline out to $z\approx 1.2$. The epoch
of cluster enrichment is yet to be identified. Most of the star
formation in the Universe occurred at $z>1$ (Madau et al. 1996;
Lanzetta et al. 2002; Dickinson et al. 2003), and fundamental plane
considerations imply that most stars in clusters formed at $z>2$ (van~Dokkum 
et al. 1998; J{\o}rgensen et al. 1999). Therefore, one expects
{\it synthesis} of most cluster metals prior to the redshifts where
they can presently be observed. However, it does not necessarily
follow that the metals are in place in the ICM at such early
epochs. As observations push back the enrichment era, scenarios where
ICM metal injection from galaxies rapidly follows their synthesis
during the early period of active star formation are favored over
those with delayed metal release from galaxies, for example, via ongoing
SN~Ia-driven winds or stripping of enriched galaxy gas halo.

\subsection{Metallicities, Abundance Patterns, and their
Interpretation}

Although often blended, emission features from O, Ne, Mg, Si, S, Ar,
Ca, Fe, and Ni were measured in {\it ASCA} spectra of individual
clusters. However, most of these are not detectable for the vast
majority of the 273 observed systems. In order to obtain abundances
with small statistical uncertainties, and smooth over both
instrumental and astrophysical systematic effects and biases,
Baumgartner et al. (2004) undertook joint analyses of
multiple observations grouped into ``stacks'' of 13--47 clusters (Baumgartner 
et al. 2004)
according to temperature (11 keV-wide bins centered on temperatures
from 0.5--10.5 keV) and metallicity (high- and low-abundance
bins). Ultimately, the two abundance stacks were merged for each
temperature, and mean Si, S, Ar, Ca, Fe, and Ni abundances proved
robust and reliable.

The results of the stacking analysis are summarized as follows (Fig. 1.1). 
For rich clusters ($kT>4$ keV), the ratio of Si to Fe is
$1.5-3$ times solar and displays an increasing trend with
temperature. However, in contrast, the S-to-Fe ratio is solar or less
(i.e. Si/S $\approx$ 3 times solar), and Ar and Ca are markedly
subsolar with respect to Fe. The Ni-to-Fe ratio is $3-4$ times solar
in these systems. This confirms and generalizes previous results
(Mushotzky et al. 1996; Fukazawa et al. 1998; Dupke \& White 2000a;
Dupke \& Arnaud 2001). Intermediate-to-poor clusters ($2~{\rm
keV}<kT<4~{\rm keV}$) have subsolar ratios of both Si and S, with
respect to Fe; however, the Si-to-S ratio remains high at about twice
solar. This trend with temperature was previously identified by
Fukazawa et al. (1998) and by Finoguenov, David, \& Ponman (2000).
The trend of higher $\alpha$/Fe ratios in the hotter clusters with
lower Fe/H (Fig. 1.2) is reminiscent of trends in Galactic stars and
suggests the existence of ICM enrichment by a stellar population that
is itself promptly enriched (i.e. by SNe~II), as inferred for
stars in bulges and elliptical galaxies.

Baumgartner et al. (2004) compare these ratios with those in various other systems,
including the Galactic thin (Fig. 1.3) and thick disks, damped
Ly$\alpha$ systems, Lyman-break galaxies, and the Ly$\alpha$
forest. In general, the high Si/Fe is not out of line with other
systems of comparable Fe/H inferred to undergo rapid and efficient
conversion of gas into stars. However, the low ratios of Ca to Fe and
S to Si, and the high ratio of Ni to Fe evidently are unique to the
ICM: a clear analog among observed classes of systems
to the population responsible for enriching the ICM is not evident.

\begin{figure*}[t]
\centering
\includegraphics[width=0.64\columnwidth,angle=90,clip]{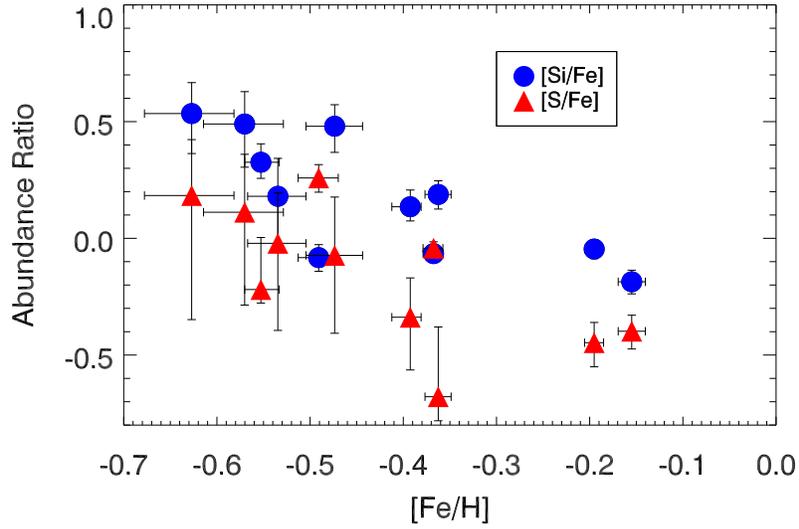}
\vskip 0pt \caption{
Si/Fe and S/Fe ratios vs. Fe/H ratio, expressed as the
logarithm with respect to solar. (From Baumgartner et al. 2004.)}
\label{si_s_vs_fe}
\end{figure*}

\vfil

\begin{figure*}[b]
\centering
\includegraphics[width=0.64\columnwidth,angle=90,clip]{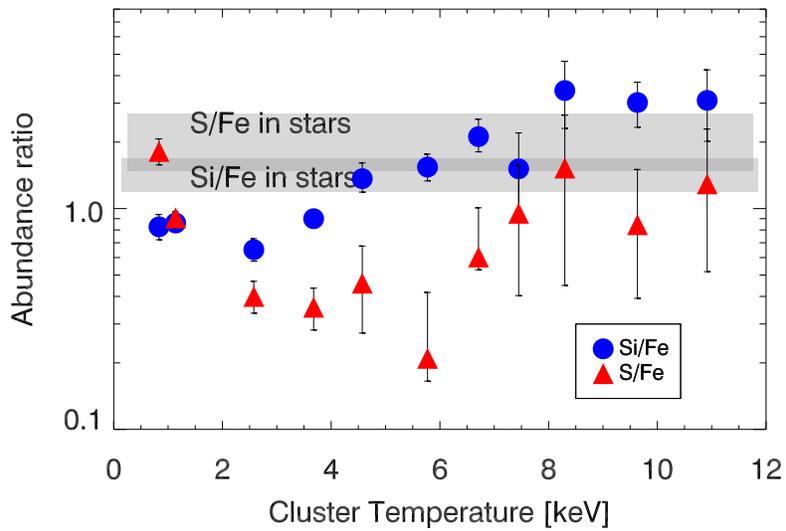}
\vskip 0pt \caption{Si/Fe and S/Fe ICM abundances vs. $kT$ compared to those
found in the Milky Way from Timmes, Woosley, \& Weaver (1995). The
range of Galactic Si/Fe is illustrated with the lower shaded
rectangle, S/Fe with the upper, overlapping rectangle (Baumgartner et al.
2004).}
\label{ratios_vs_MW}
\end{figure*}

\clearpage

\section{On the ICM Abundance Anomalies}

Since bulges account for $\sim 90$\% of the stellar mass in clusters,
it is straightforward to estimate ICM enrichment under standard
assumptions. Consider a simple stellar population with a local IMF
(e.g., Kroupa 2001), assume that stars more massive than 8 $M_{\odot}$
explode as SNe~II, that the rate of SN~Ia explosions is as determined in
nearby elliptical galaxies (0.16 SNU, where 1 SNU $\equiv$ 1 SN per
$10^{10}$ solar blue luminosities per century; Cappellaro, Evans, \&
Turatto 1999) and injected for a duration of $10^{10}$ yr, and utilize
SNe~II and SNe~Ia (``W7'' model) yields from Nomoto et al. (1997a, b) with
the former averaged over a standard IMF extending to 40 $M_{\odot}$ as
in Thielemann, Nomoto, \& Hashimoto (1996). The fractions of Fe
originating in SNe~Ia separately inferred from the Si/Fe and S/Fe ratios
are shown in Figure 1.4, illustrating that {\it no} combination of
standard SNe~Ia and SNe~II can explain the ICM abundance pattern. Baumgartner 
et al. (2004) 
demonstrate that this holds regardless of which published SN~II yields
are considered (see also Gibson, Loewenstein, \& Mushotzky 1997;
Loewenstein 2001).

\begin{figure*}
\centering
\includegraphics[width=0.77\columnwidth,angle=90,clip]{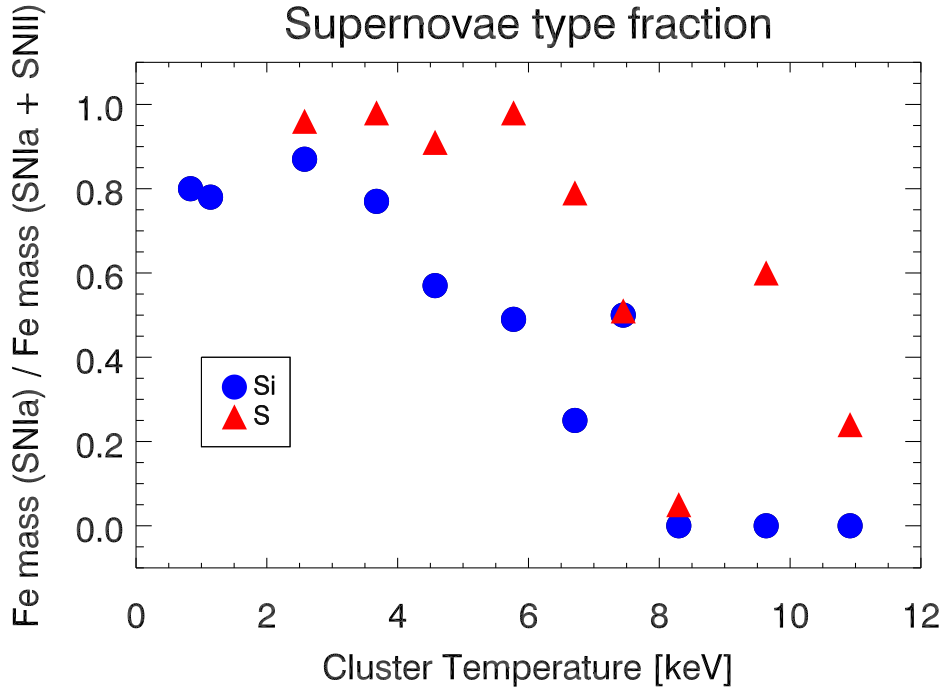}
\vskip 0pt \caption{
Decomposition into SNe~Ia and SNe~II, using standard yields and
the observed Si/Fe and S/Fe ratios, expressed in terms of the fraction
of Fe from SNe~Ia. (From Baumgartner et al. 2004.)}
\label{sn_yields_fractions}
\end{figure*}

In addition to ratios, it is instructive to consider the ICM mass in
metals, normalized to the total starlight in cluster galaxies. Table
1.2 compares the mass-to-light ratios of the elements addressed in the
stacking analysis, computed using the standard assumptions detailed
above, with the observed averages in three ICM temperature
regimes. The predicted SN~II and SN~Ia contributions are shown
separately, as well as their sum. For comparison, the predictions for
a delayed detonation SN~Ia model (``WDD1,'' Nomoto et al. 1997b) are
shown along with those using the standard W7 model; use of the former
generally exacerbates the problems detailed below.

\begin{table*}
\refsize
\caption{Metal Mass-to-Light Ratios}
\begin{tabular}{ccccccccc}
\hline\hline
& \multicolumn{4}{c}{Predicted} & & \multicolumn{3}{c}{Observed} \\
\cline{2-5} \cline{7-9} \\
&SNe~II  & SNe~Ia & SNe~Ia &SNe~II  && $kT>8$ & $4<kT<8$ & $2<kT<4$\\
&      & W7   & WDD1 & W7   && (keV) & (keV) & (keV) \\ \hline
Si & $9.3\times 10^{-3}$ & $2.4\times 10^{-4}$ & $5.6\times 10^{-4}$ & $9.3\times 10^{-3}$ &
   & \, \ \ $2.3\times 10^{-2}$ & \, \ \ $1.3\times 10^{-2}$ & $7.5\times 10^{-3}$\\
S  & $3.1\times 10^{-3}$ & $1.4\times 10^{-4}$ & $3.4\times 10^{-4}$ & $3.3\times 10^{-3}$ &
   & \, \ \ $4.5\times 10^{-3}$ & \, \ \ $2.5\times 10^{-3}$ & $2.3\times 10^{-3}$\\
Ar & $6.1\times 10^{-4}$ & $2.6\times 10^{-4}$ & $6.9\times 10^{-5}$ & $6.3\times 10^{-4}$ &
   & $<9.6\times 10^{-4}$ & $<5.5\times 10^{-4}$ & ... \\
Ca & $4.5\times 10^{-4}$ & $2.0\times 10^{-4}$ & $6.5\times 10^{-5}$ & $4.7\times 10^{-4}$ &
   & $<6.6\times 10^{-4}$ & $<3.8\times 10^{-4}$ & ... \\
Fe & $6.9\times 10^{-3}$ & $1.2\times 10^{-3}$ & $8.7\times 10^{-4}$ & $8.1\times 10^{-3}$ &
   & \, \ \ $1.4\times 10^{-2}$ & \, \ \ $1.5\times 10^{-2}$ & $1.7\times 10^{-2}$\\
Ni & $4.5\times 10^{-4}$ & $2.3\times 10^{-4}$ & $5.7\times 10^{-5}$ & $6.8\times 10^{-4}$ &
   & \, \ \ $3.1\times 10^{-3}$ & \, \ \ $2.7\times 10^{-3}$ & ... \\ \hline\hline
\end{tabular}
\label{table 1.2}
\end{table*}

There are a number of notable discrepancies. Both Fe and Ni are
generally underpredicted, as is Si at high temperatures. One could
achieve reconciliation for Fe and Si by roughly doubling the numbers
of SNe~Ia and SNe~II, alterations that are not unreasonable since the SN~Ia
rate may be greater in the past and the IMF may be top-heavy (flat or
bimodal; e.g., Loewenstein \& Mushotzky 1996). One could then, perhaps,
explain the Si/Fe trend with ICM temperature as selective mass loss of
SN~II products in systems with shallower gravitational potential wells.
However, such a scenario overpredicts S, Ar, and Ca and underpredicts
Ni.

The large observed amounts of Ni require a SN~Ia rate, averaged over a
Hubble time, of more than 10 times the estimated local rate, with a
concomitant decrease in the number of SNe~II to avoid overproducing
Fe. However, this underpredicts Si. The problem is intractable using
standard yields, unless a significant additional source of metals is
considered.

What are some possible resolutions to this puzzle? The only
explanation for the high Ni abundance is a high time-averaged SN~Ia
rate and/or a high SN~Ia Ni yield---the rate must exceed
$\sim 2(0.14/y_{\rm Ia,Ni})$ SNU, where $y_{\rm Ia,Ni}$ is the SN~Ia Ni
yield in solar masses. To account for those systems with high Si
abundances, one could assume that SN~II Si yields are twice those in
the core collapse nucleosynthesis calculations currently in the
literature. Alternatively, the Si overabundance could be explained if
the number of massive SN~II progenitor stars were approximately twice
what a standard IMF predicts; but, in this case the S, Ar, Ca, and,
perhaps, Fe yields require {\it ad hoc} reduction to avoid their
overproduction.

If published yields are not grossly in error, one must appeal to an
additional source of enrichment that preferentially synthesizes Si.
These progenitors must be more plentiful, or their products more
efficiently retained, in more massive clusters. An extensive search of
the literature reveals several instances of SNe with the
desired yield-pattern, including core collapse SN from very massive
(70 $M_{\odot}$) stars (Thielemann et al. 1996), massive ($\sim
30$ $M_{\odot}$) metal-poor ($\sim 0.01$ solar) stars (Woosley \&
Weaver 1995), or supermassive (with 70 $M_{\odot}$ He cores)
metal-free stars that explode as hypernovae (Heger \& Woosley
2002). While it is unlikely that a monolithic contribution from any of
these objects truly explains the ICM abundance anomalies, their
existence demonstrates that there are nucleosynthetic channels---likely 
associated with very massive and/or very metal-poor stars---that 
result in relative enhancements of Si. Figure 1.5 shows the
respective contributions of SNe~Ia, SNe~II, and hypernovae (Heger \&
Woosley 2002) to each element in a toy model that reproduces a typical
rich cluster ICM abundance pattern (Baumgartner et al. 2004).

There are now possible signatures of Population III and hypernovae in
the level and pattern of enrichment in low-metallicity Milky Way stars
(Umeda \& Nomoto 2003), in the level and epoch of intergalactic medium 
enrichment as inferred from the Ly$\alpha$ forest (Songaila 2001), and in the
reionization of the Universe (e.g., Wyithe \& Loeb 2003). The number
of hypernovae required in the ICM is reasonable in terms of the
expected number of Population III progenitors (Loewenstein 2001).

\begin{figure*}
\centering
\includegraphics[clip,height=1.00\columnwidth,angle=90]{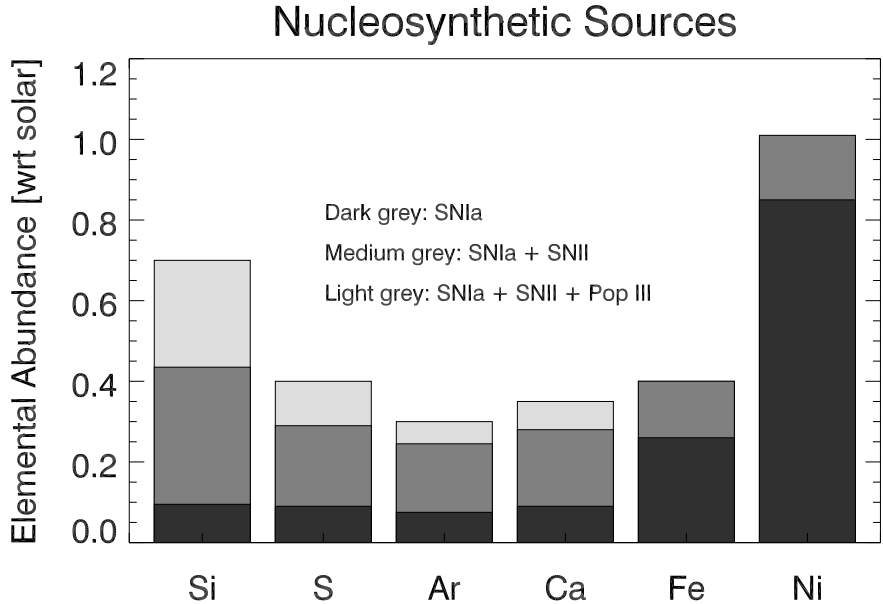}
\vskip 0pt \caption{
Relative contributions to the enrichment of various elements
in a typical clusters for a toy model including Population III
hypernovae in addition to SNe~Ia and SNe~II in standard numbers with
standard yields (Baumgartner et al. 2004).}
\label{decomp}
\end{figure*}

\section{{\it Chandra} and {\it XMM-Newton} Results on Abundance
Gradients}

{\it ASCA} data revealed clear evidence of central excesses of Fe in
clusters with cooling flows and in clusters not involved in recent major
mergers (Matsumoto et al. 1996; Ezawa et al. 1997; Ikebe et al. 1997, 1999; 
Xu et al. 1997; Kikuchi et al. 1999; Dupke \& White 2000a, b), which are
consistent with generally flat metallicity
profiles outside of the central $\sim 2'$ (Tamura et al. 1996; Irwin,
Bregman, \& Evrard 1999; White 2000). Limitations in {\it ASCA}
imaging precluded resolving gradients in Fe or drawing robust
conclusions on the spatial distribution of $\alpha$/Fe ratios, although 
there were indications that $\alpha$ elements may not display
the central excess (Finoguenov et al. 2000; Fukazawa et al. 2000; Allen et al. 
2001). {\it BeppoSAX} provided Fe profiles with better
spatial resolution, and De Grandi \& Molendi (2001) attributed the
central Fe excess to injection by SNe~Ia associated with the stellar
population in the central cluster galaxy. If this is the case, milder
or absent central excesses in SN~II-synthesized $\alpha$ elements are
expected, as claimed by Finoguenov et al. (2000).

These questions of gradients in abundance ratios are addressed through
observations with the {\it Chandra} (Weisskopf et al. 2002) and {\it
XMM-Newton} (Jansen et al. 2001) X-ray observatories. The former
represents a huge leap in broad-bandpass imaging, with an angular
resolution $\sim 0.''5$, the latter in collecting area (a total of
$\sim 5$ times that of {\it ASCA}, with an order of magnitude better
angular resolution). New results demonstrate that the central Fe
excess is concentrated to within 100 kpc of the cluster center
(Iwasawa et al. 2001; Kaastra et al. 2001; Lewis, Stocke, \& Buote 2002; Smith 
et al. 2002), that there may be a central metallicity
inversion (Johnstone et al. 2002; Sanders \& Fabian 2002; Schmidt,
Fabian, \& Sanders 2002; Blanton, Sarazin, \& McNamara 2003; Dupke \& White 
2003), and that abundance profiles are generally very flat
from $\sim 100$ kpc out to a significant fraction of the virial radius
(Schmidt, Allen, \& Fabian 2001; Tamura et al. 2001). The apparent
inversion may be an artifact of using an oversimplified model in
spectral fitting (Molendi \& Gastaldello 2001). The question of the
presence of $\alpha$/Fe gradients is not yet definitively settled
(David et al. 2001), although early results indicate a mix of
abundances in the central regions that is closer to the SN~Ia pattern
than is globally the case (Tamura et al. 2001; Ettori et al. 2002).

\section{{\it XMM-Newton} RGS Measurements of CNO}

The {\it XMM-Newton} Reflection Grating Spectrometer provides high-spectral 
resolution ($E/\Delta E$ ranges from 200 to 800 over the
0.35--2.5 keV RGS bandpass; den~Herder et al. 2003) spectroscopy in the
wavelength region containing the strongest X-ray features of the
elements carbon through sulfur. Since individual lines are more
distinctly resolved than in previous studies using CCD spectra (where
$E/\Delta E \approx  50$), abundance determinations are much less sensitive
to the assumed thermal emission model. C, N, and O lines are all
widely accessible for the first time. These are unique probes of star
formation and the era of ICM enrichment because of their distinctive
nucleosynthetic origins. For an extended X-ray source such as a galaxy
cluster, RGS spectra correspond to an emission-weighted average over
the inner $\sim 1'$ (see Peterson et al. 2003)---within the central
galaxy (if there is one) for a nearby cluster; thus, caution must
be exercised in applying the results to the source as a whole.

\subsection{Oxygen}

The strong 0.65 keV OV~{\sc iii} Ly$\alpha$ feature is measured with {\it
XMM-Newton} in clusters out to redshifts as high as 0.3. The O/Fe
ratio for the sample of cooling flow clusters in Peterson et
al. (2003) is shown in Figure 1.6. [Note that Peterson et al. adopt
the definition of solar abundances from Anders \& Grevesse (1989), where
log~(Fe/H) + 12 = 7.67 and log~(O/H) + 12 = 8.93, $\sim 1.5$ and $\sim 1.7$ 
times higher than currently more commonly used ``cosmic
abundances'' of Fe and O, respectively.]  O/Fe is generally subsolar,
and overlaps with the range measured in Galactic stars of comparable
Fe/H (Reddy et al. 2003). Low O abundances were previously reported in
Abell 1795 (Tamura et al. 2001a), Abell 1835 (Peterson et al. 2001),
S\`ersic 159-03 (Kaastra et al. 2001), Abell 496 (Tamura et
al. 2001b), NGC 5044, the central (elliptical) galaxy in the WP23
group (Tamura et al. 2003), and the elliptical galaxy NGC 4636 (Xu et
al. 2002), which resembles a group in X-rays. The Galactic O/Fe-Fe/H
trend is thought to reflect the delayed enrichment in Fe from SNe~Ia
relative to that of O from short-lived, massive SN~II progenitors. In
the intergalactic medium of clusters and groups, we may be seeing
systematic variations in the relative stellar lock-up fractions of
SN~II and SN~Ia products, relative retention of injected SN~II and SN~Ia
material, or IMF. Low O/Si ratios in these systems were interpreted as
a possible signature of enrichment by hypernovae associated with
Population III (Loewenstein 2001). Intriguingly, as shown in Figure
1.7, Mg appears not be a good surrogate for O, as often assumed in
studies of elliptical galaxies.
 
\begin{figure*}
\centering
\includegraphics[clip,height=0.85\columnwidth,angle=0]{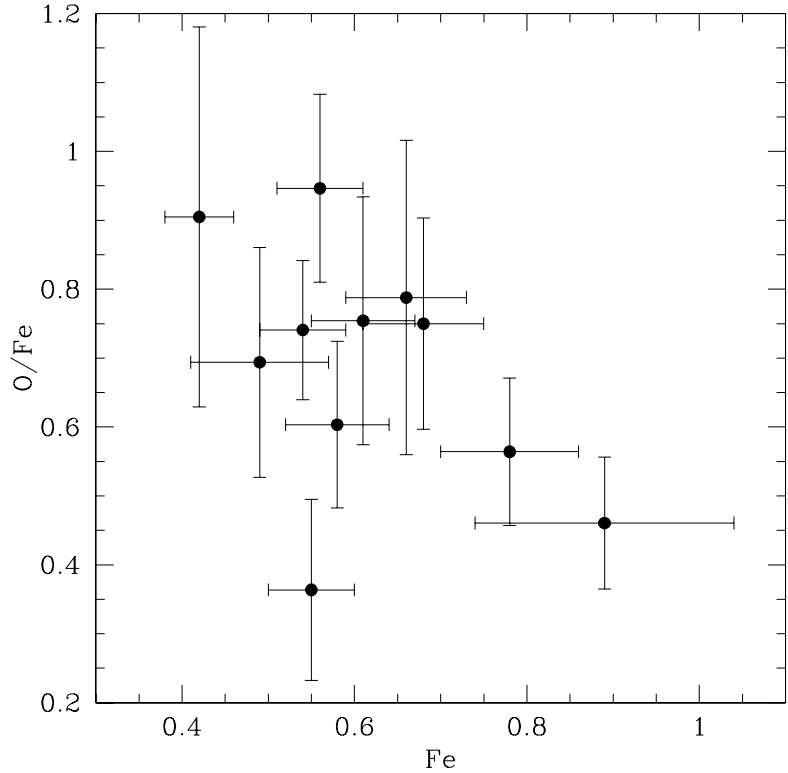}
\vskip 0pt \caption{
O/Fe vs. Fe for cooling flow clusters.  (Adapted from Peterson et al. 2003.)}
\label{o-fe}
\end{figure*}

\begin{figure*}
\centering
\includegraphics[clip,width=.48\columnwidth,angle=0]{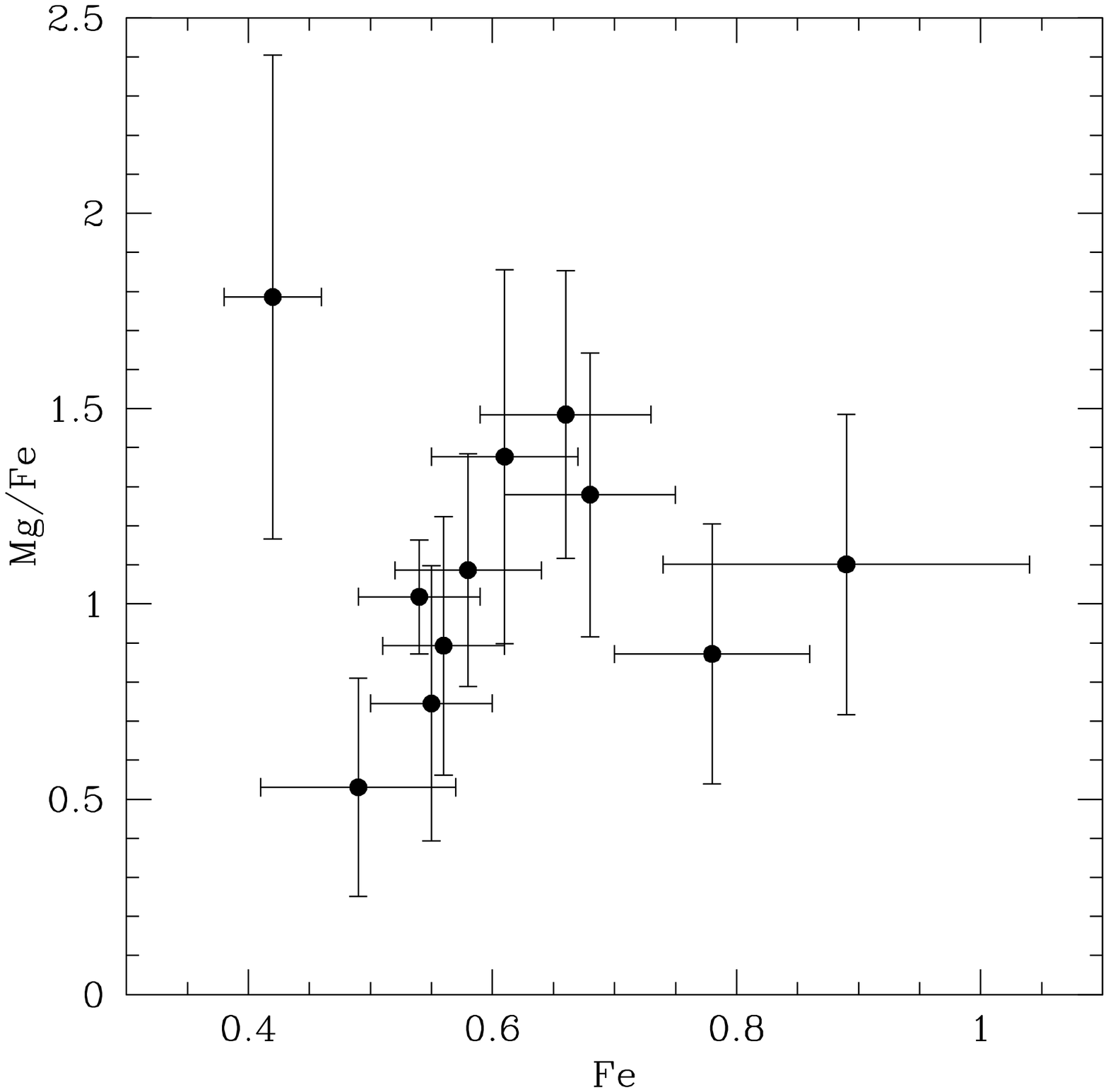} \hfil
\includegraphics[clip,width=.48\columnwidth,angle=0]{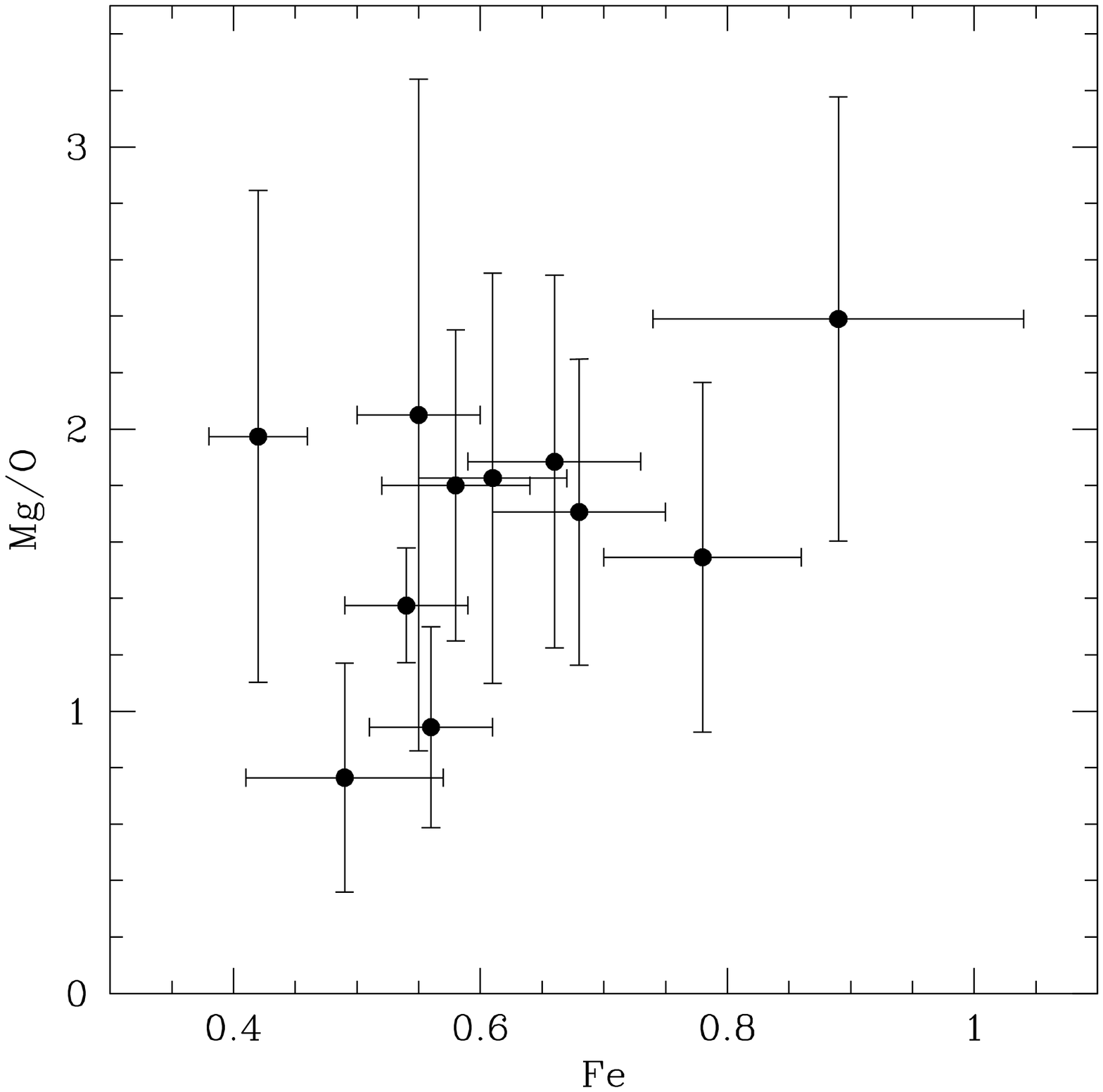}
\vskip 0pt \caption{
Mg/Fe and Mg/O vs. Fe for cooling flow clusters. (From Peterson et al. 2003.)}
\label{mg-fe-o}
\end{figure*}

\subsection{Carbon and Nitrogen}

The Ly$\alpha$ lines of the H-like ions of C and N are weak compared
to O, and fall in a less sensitive and less well-calibrated wavelength
region of the RGS. As a result, there presently are few measurements
of C and N: C and N in A 496 and M87 (Peterson et al. 2003) and N
in the elliptical galaxies NGC 4636 (Xu et al. 2002) and NGC 5044
(Tamura et al. 2003). N is of particular interest, since its origin in
intermediate-mass stars provides crucial leverage in distinguishing
the mass function of the enriching stellar population. In these
objects C (when measured) and N are $\sim$solar, and C/O and N/O are
$\sim$twice solar [solar as in Anders \& Grevesse 1989:
loG(C/H) + 12 = 8.56, log~(N/H) + 12 = 8.05]. On a plot of N/O vs. O/H,
these systems overlap with measurements in extragalactic H~{\sc ii} regions
(Pettini et al. 2002), where the secondary production of N dominates,
but do extend to somewhat higher N/O at the same O/H.

\section{{\it XMM-Newton} Observations of M87}

Because of its brightness, M87 in the Virgo cluster is among the most
extensively X-ray spectroscopically studied extended extragalactic
objects. Its proximity enables one to extract detailed plasma
conditions and composition, and to evaluate and correct for systematic
effects. Previous work based on spectra from the {\it Einstein}
(Canizares et al. 1982; Stewart et al. 1984; White et al. 1991; Tsai
1994a, b), {\it ROSAT} (Nulsen \& Bohringer 1995; Buote 2001), and {\it
ASCA} (Matsumoto et al. 1996; Hwang et al. 1997; Buote, Canizares, \&
Fabian 1999; Finoguenov \& Jones 2000; Shibata et al. 2001)
observatories is rapidly becoming superseded by {\it Chandra} and {\it
XMM-Newton} data.

The RGS results for M87 show that, in the inner $\sim 10$ kpc,
abundances of C, N, Ne, and Mg have best-fit values (relative to
Anders \& Grevesse 1989) of 0.7--1.0 solar, with 90\% uncertainties of
0.2--0.3, while Fe is $0.77\pm 0.04$ solar, and O $0.49\pm 0.04$ solar
(Sakelliou et al. 2002).  With respect to the M87 stellar
abundances (Milone, Barbuy, \& Schiavon 2000), the $\alpha$ elements are
apparently preferentially diluted with respect to Fe.

Gastaldello \& Molendi (2002) derive detailed abundance distributions
from analysis of {\it XMM-Newton} EPIC-MOS CCD data, obtaining O, Ne,
Mg, Si, Ar, Ca, Fe, and Ni in 10 radial bins extending to $14'$. The
MOS results are generally consistent with the RGS and indicate
negative gradients on arcminute scales in all of the elements, with
the possible exception of Ne. Gradients in $\alpha$/Fe ratios are
modest. Ratios with respect to Fe vary among the $\alpha$ elements:
O/Fe, Ne/Fe, Mg/Fe, and S/Fe are subsolar; Si/Fe, Ar/Fe, and Ca/Fe are
slightly supersolar. Ni/Fe increases from $\sim 1.5$ solar in the
center to $\sim 2.5$ solar at 50 kpc. Not surprisingly, Gastaldello \&
Molendi find that no unique combination of published SN~Ia (either
deflagration or delayed detonation models) and SN~II yields is
consistent with the full abundance pattern at any radius. Moreover,
the combination of SN subtypes that provides the best fit varies with
radius (i.e. delayed detonation models provide a better fit in
the center, W7 further out where Ni/Fe is greater).

Finoguenov et al. (2002) utilize a cruder spatial division, thus
obtaining higher precision for individual abundances (this is enhanced
by their inclusion of {\it XMM-Newton} EPIC-PN CCD data). They propose
explaining the observed abundance pattern and its radial variation
with (1) a radially increasing SN~II/SN~Ia ratio, (2) high Si and S
yields from SNe~Ia (favoring delayed detonation models) and an {\it ad
hoc} reduction in SN~II S yields, and (3) a radial variation in SN~Ia
yields (corresponding to delayed detonation models with different
deflagration-to-detonation transition densities). As they note,
deflagration models must also play a role to account for the high
Ni/Fe ratio found in other clusters. Whether such a complex model can
coherently explain cluster abundance patterns in general is
unclear. Matsushita, Finoguenov, \& Bohringer (2003) follow this up
with more sophisticated thermal modeling, and an analysis that
includes individual line ratios, and they reach conclusions that are
qualitatively similar in terms of the outwardly increasing importance
of SNe~II, the diversity of SNe~Ia, and the yields of Si and S.

It is important to keep in mind that these new results are for the
central $\sim 70$ kpc---still well within the influence of the M87
galaxy. The larger-scale cluster plasma (as studied in the ACC project
discussed above) is not as profoundly affected by SNe~Ia, and, indeed,
there are indications of a transition to a more cluster-like abundance
pattern at large radii in M87 (Finoguenov et al. 2002). In fact, the
narrowing redshift window when the Fe in clusters must be synthesized
and mixed into the ICM, as inferred from the lack of Fe abundance
evolution, is becoming comparable to the time delay characteristic of
many models of SN~Ia explosions.

\section{Concluding Remarks}

The era of high-precision X-ray spectroscopy has arrived. Accurate
and robust measurements can now be made of abundances and their
variation in space and time, for a wide variety of elements in
intracluster and intragroup media. This enables astronomers to utilize
the ICM, where metals synthesized in cluster galaxy stars over the age
of the Universe accumulated, as a laboratory for testing theories of
the environmental dependence and impact of star formation.

\subsection{Mean ICM Metallicities and Redshift Dependence}

There are now hundreds of measurements, out to redshifts greater than
1, of Fe abundances and many tens of measurements (or significant
upper limits) of Si, S, Ar, Ca, and Ni. The distribution function of
Fe abundance outside of the central regions in rich clusters is
sharply peaked around $\sim 0.4$ solar, and shows no evidence of a
decline with increasing redshift out to $z\approx 1$. With the average Fe
yield per current mass of stars $\sim 3$ times solar, a prodigious
rapid enrichment is implied: a time average of $\sim 0.1$ $M_{\odot}$
yr$^{-1}$ of Fe per (present-day) $L_*$ galaxy integrated over 2
Gyr. This is equivalent to the nucleosynthetic yield of $\sim 1$ SN~II
yr$^{-1}$ per $L_*$ galaxy, corresponding to $\sim 100$ $M_{\odot}$
yr$^{-1}$ per $L_*$ galaxy forming stars with a Salpeter IMF, or $\sim
20$ SNU of SNe~Ia! As a point of reference, note that the mean star
formation rate in the Universe at high redshift, normalized to the
cluster light, is $\sim 0.02$ $M_{\odot}$ yr$^{-1}$ per $L_*$ galaxy.
So, despite a stellar fraction consistent with the universal average,
several times the amount of metals that might be expected was
synthesized in clusters. It is intriguing that Lyman-break galaxies
were enriched to similar levels on a comparably brief time interval.
In the extreme overdense regions that represent clusters of galaxies,
the primordial IMF was skewed toward high masses, or an additional
separate primordial population of massive stars was present. Clusters
are thus strong candidates for the source of the missing metals at
$z\approx 3$ (Finoguenov et al. 2003), even though only about 5\% of
the total universal baryon content resides in clusters.
 
\subsection{Abundance Gradients}

Centrally concentrated ($<100$ kpc) excesses in Fe are common, and
evidently exclusively occur in those clusters that can be
characterized either as cooling flow clusters, as clusters with a
massive central galaxy, or as clusters without recent major merging
activity. An explanation of this phenomenon awaits additional {\it
Chandra} and {\it XMM-Newton} results on the radial distribution of
$\alpha$ elements. Early indications are that SNe~Ia play a more
important role at the location of central cluster galaxies, although
other giant elliptical galaxies evidently did not retain such large
masses of interstellar Fe. Metallicity gradients evidently are mild or
absent beyond these central regions.
 
\subsection{Abundance Patterns}

There is more Si, Fe, and Ni than one might expect based on the
observed stellar population. The relative abundances of elements
measured in the ICM show clear departures from solar ratios and the
abundance pattern in the Galactic disk and other well-studied galactic
and protogalactic systems. The ICM abundance pattern shows systematic
variations with ICM temperature (i.e. cluster gravitational
potential well depth). These variations are apparent even in the
relative abundances, which are not generally in solar ratios, among
different $\alpha$ elements (O, Mg, Si, S, Ar, and Ca). Contribution from
an additional, primordial source of metals may be required to finally
explain these anomalies. The ICM Ni/Fe ratio is $2-4$ times solar,
higher than in any other known class of object where this ratio is
measured. Since both Fe and Ni are efficiently synthesized in SNe~Ia,
and given the importance of SNe~Ia as fundamental probes of the
cosmological world model, the origin of the Ni excess is clearly
worthy of further investigation.

\subsection{The Future}

As the {\it Chandra} and {\it XMM-Newton} cluster databases mature,
one can expect many more measurements of abundance gradients (of
$\alpha$ elements, as well as Fe), tight constraints on the evolution
of Si and O abundances out to $z>0.4$ and Fe out to $z>1$, and
additional accurate measurements of C and N. {\it ASTRO-E2}, 
scheduled for launch in 2005, will provide true spatially resolved,
high-resolution X-ray spectroscopy, which will yield cleaner measurements of
N, O, Mg, and Ne, and their gradients.

\vspace{0.3cm}
{\bf Acknowledgements}.
I would like to acknowledge the
organizers for their kind invitation to participate in this meeting,
and for all of their efforts in assembling the conference and these
proceedings. My gratitude extends to Richard Mushotzky and Wayne
Baumgartner for their assistance in preparing this review.

\def\aa{{A\&A}}
\def\aas{{A\&AS}}
\def\aj{{AJ}}
\def\annrev{{ARA\&A}}
\def\apj{{ApJ}}
\def\apjs{{ApJS}}
\def\baas{{BAAS}}
\def\mnras{{MNRAS}}
\def\nat{{Nature}}
\def\pasp{{PASP}}
\def\pasj{{PASJ}}

\end{document}